\documentclass[pra,preprint,final,nototal,showpacs,superscriptaddress,fleqn,amsmath,amsfonts,tightenlines,noindent]{revtex4}
\usepackage{graphicx}
\usepackage{subfigure}
\usepackage{enumerate}
\newcommand{\ket}[1]{\mid \! #1 \rangle}

\newcommand{\figbox}[1]{%
  \fbox{%
    \vbox to 1in{%
    \vfil
    \hbox to 2in{%
      \hfil
      #1%
      \hfil}%
    \vfil}}}
                                                                                
\newcommand{\goodgap}{%
  \hspace{\subfigcapskip}}

\begin{document} 
\title{Ultracold atoms in optical lattices}

\author{D.B.M. Dickerscheid}
\affiliation{Institute for Theoretical Physics, University of Utrecht, 
Princetonplein 5, 3584 CC Utrecht, The Netherlands}
\affiliation{Lorentz Institute, Leiden University, P. O. Box 9506, 
2300 RA Leiden, The Netherlands}

\author{ D. van Oosten}
\affiliation{Institute for Theoretical Physics, University of Utrecht, 
Princetonplein 5, 3584 CC Utrecht, The Netherlands}
\affiliation{Debye Institute, University of Utrecht, 
Princetonplein 5, 3584 CC Utrecht, The Netherlands}

\author{ P.J.H. Denteneer}
\affiliation{Lorentz Institute, Leiden University, P. O. Box 9506, 
2300 RA Leiden, The Netherlands}

\author{ H.T.C. Stoof} 
\affiliation{Institute for Theoretical Physics, University of Utrecht, 
Princetonplein 5, 3584 CC Utrecht, The Netherlands}

\begin{abstract} 
Bosonic atoms trapped in an optical lattice at very low temperatures, can 
be modeled by the Bose-Hubbard model. In this paper, we propose a slave-boson approach 
for dealing with the Bose-Hubbard model, which enables us to 
analytically describe the physics of this model at nonzero temperatures.
With our approach the phase diagram for this model at nonzero temperatures can be quantified.
\end{abstract}
 \pacs{03.75.-b,67.40.-w,39.25.+k}
 \maketitle

\section{Introduction}

\begin{figure} \includegraphics[width=10cm]{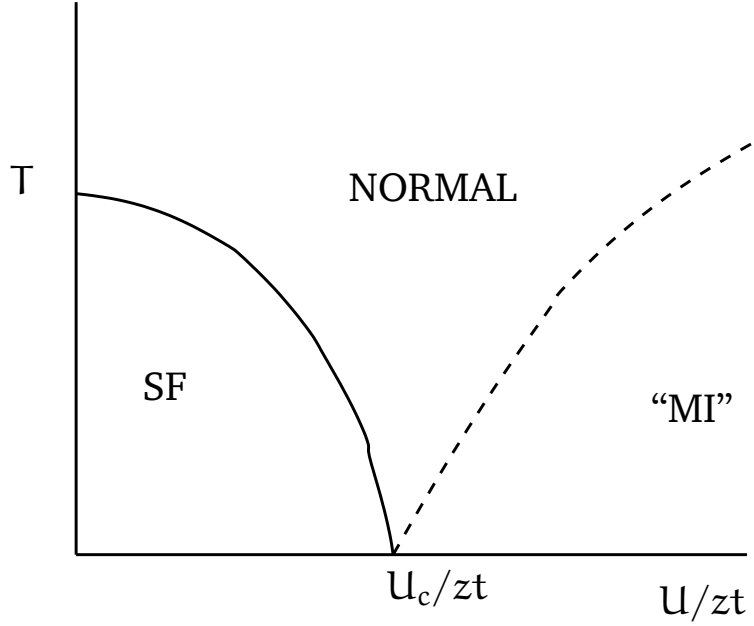} \caption{Qualitative phase diagram for a fixed
and integer filling fraction in terms of the temperature $T$ and the dimensionless coupling constant $U/z t $, with superfluid (SF),
normal and Mott insulating phases (MI). Only at $T=0$ a true Mott insulator exists.  }\label{qualphasediag}
\end{figure}

\begin{figure} \includegraphics[width=10cm]{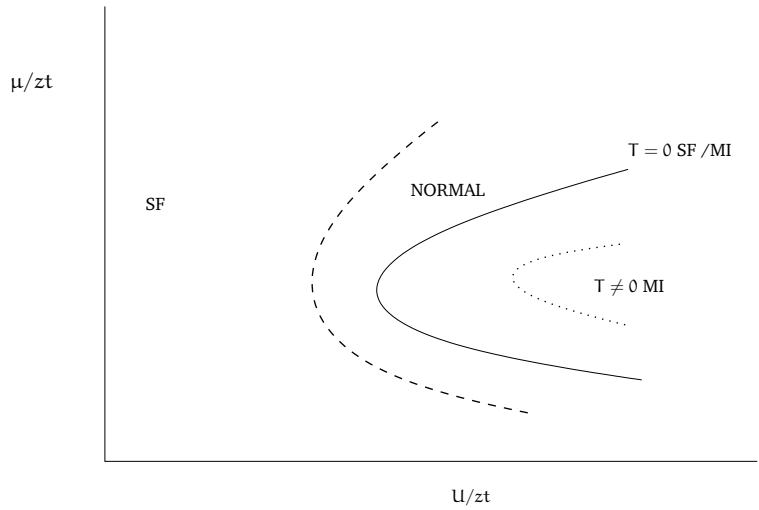} \caption{Qualitative phase diagram in
terms of the chemical potential $\mu/zt $ and the dimensionless  coupling constant $U/zt$. For nonzero temperatures a normal phase
appears.}\label{schemnonzero} \end{figure}

The physics of the Bose-Hubbard model was the subject of intensive study for some years after
the seminal paper by Fisher \emph{et al.}, which focused on the behavior
of bosons in a disordered environment \cite{Fisher1989}.
More recently  it has been 
realized 
that the Bose-Hubbard model can also be applied to bosons trapped in so-called optical lattices
\cite{Jaksch1998}, and mean-field theories \cite{Sheshadri1993,Oosten2001,vanOosten2003a}
and exact diagonalization \cite{Roth2003}  have been succesfully applied to these systems in 
one, two and three dimensional systems.
The experiments performed by Greiner \emph{et al.} \cite{Greiner2002} have 
confirmed
the theoretically predicted quantum phase transition, i.e., a phase transition induced by quantum fluctuations,
between a superfluid and a Mott-insulating phase. A review of the work carried out 
 in this field has been given by Zwerger \cite{Zwerger2002}.
Strictly speaking the above mentioned quantum phase transition 
occurs only at zero temperature \cite{Sachdev}.
 At nonzero temperatures there is  a `classical' phase transition, i.e., a phase transition induced by thermal 
fluctuations, between a superfluid phase  and a normal phase and there is only a crossover between the normal phase 
and a Mott insulator.  It is important to mention here that a
 Mott insulator is by definition  incompressible.
In principle there  exists, therefore, no Mott insulator 
for any nonzero temperature
where we always have a nonvanishing compressibillity. Nevertheless,
there is a region in the phase diagram  where the compressibillity is very close to 
zero and it is therefore justified to
call this region  for all practical purposes a Mott insulator \cite{vanOosten2003a}.
Qualitatively this phase diagram  
is sketched in Fig. \ref{qualphasediag} for a fixed density. 
This figure shows how at a sufficiently small but
 nonzero temperature we start with a superfluid for small positive on-site interaction 
$U$,  we encounter a phase transition to a normal phase 
as the interaction strength increases,
and ultimately crossover to a Mott insulator
for even higher values of the interaction strength.
We can also incorporate this nonzero 
temperature behaviour into  the phase diagram in Fig. \ref{schemnonzero}. 
This figure shows how at zero temperature we only have a superfluid and a Mott insulator phase, but as 
the temperature is increased a normal phase appears in between these two phases.

The aim of this paper is to extend the mean-field approach for the Bose-Hubbard model to include nonzero temperature
effects and make the qualitative phase diagrams in Figs. \ref{qualphasediag} and \ref{schemnonzero} more 
quantitative.
To do that we make use of auxiliary particles that are known as slave bosons. 
The idea behind this is that if we consider a single lattice site, the occupation number on that 
site can be any integer. 
With each different occupation number we identify  a new particle. 
Although this means that we introduce a lot of different 
new particles, the advantage of this procedure is that it allows us to transform the on-site repulsion 
into an energy contribution that is quadratic in terms of the new particles.
Because we want to be able to uniquely label each different state of the system,
the new particles cannot independently be present at each lattice site. 
That is why we have to introduce a constraint.
Using this we derive within a functional-integral 
formalism  an effective action for the superfluid order parameter which depends on the temperature.
The equivalence with previous work at zero temperature is demonstrated.

The outline of the paper is as follows.  In Sec. \ref{theory} we introduce the slave-boson formalism 
and derive an 
effective action for the superfluid order parameter. In Sec. \ref{meanfieldresults} we present the 
zero and nonzero temperature mean-field results. 
The remainder of the paper is devoted to the effect that the creation  of 
quasiparticle-quasihole pairs have on the system.

\section{Slave-boson theory for the Bose-Hubbard model}\label{theory}

In this section we formalize the above introduced idea of the slave bosons. We rewrite the Bose-Hubbard
model in terms of these slave bosons within a path-integral formulation and derive an effective action for the 
superfluid order parameter, which then describes all the physics of our Bose gas in the optical lattice.

The slave-boson technique was introduced by Kotliar and Ruckenstein \cite{Kotliar1986}, who used it 
to deal with the fermionic Hubbard model.
A functional integral approach to the problem of hard-core bosons hopping on a lattice 
has been previously put forward by Ziegler \cite{Ziegler1993} and Fr\'esard \cite{Fresard1994}.
Let us first shed some light on this slave-boson formalism.
We consider a single site of our lattice. If the creation and anihilation
operators for the bosons are denoted by $\hat{a}^{\dagger}_{i}$ and
$\hat{a}^{\phantom \dagger}_{i}$ respectively, we can form the number operator
$\hat{N}_{i} =\hat{a}^{\dagger}_{i} \hat{a}^{\phantom \dagger}_{i} $, which counts the number of bosons at 
the site $i$. In the slave-boson formalism, for any occupation number a pair of
 bosonic creation and annihilation operators is introduced that 
create and annihilate the state with precisely that given integer number of particles.
The original occupation number states $\ket{n_{i}}$ are now decomposed as 
$\ket{n^{0}_{i},n^{1}_{i}, \ldots}$, where 
$n^{\alpha}_{i}$ is the eigenvalue of the number operator 
$\hat{n}^{\alpha}_{i} \equiv (\hat{a}_{i}^{\alpha})^{\dagger} \hat{a}_{i}^{\alpha}$
formed by the pair of 
creation $(\hat{a}_{i}^{\alpha})^{\dagger}$ and annihilation 
$\hat{a}_{i}^{\alpha}$ operators that 
create and annihilate bosons of type $\alpha$ at the site $i$.
As it stands, this decomposition is certainly not unique. 
For example, the original state $\ket{2}$ could be written as $\ket{0,0,1,0,\ldots}$
or as $\ket{0,2,0,\ldots }$. Our Hilbert space thus greatly increases.
To make sure that every occupation occurs only once we have to introduce an additional  
constraint, namely
\begin{equation}\label{firstconstraint}
\sum_{\alpha} \hat{n}^{\alpha}_{j}
= 1
\end{equation} for every site $j$.
This constraint thus makes sure that there is always just one slave boson per site.
Because in the positive $U$ Bose-Hubbard model bosons on the same site repel each other,  high on-site occupation 
numbers are  disfavored. It is therefore conceivable that a good approximation of the physics of the Bose-Hubbard 
model is obtained by  allowing a relatively small maximum number,  e.g. two or three or four, of bosons per site.

As is well known, the Hamiltonian of the Bose-Hubbard model reads,
\begin{equation}\label{sep2}
\hat{H} = - \sum_{\langle i,j \rangle}
\hat{a}^{\dagger}_{i} t^{\phantom \dagger}_{ij} \hat{a}^{\phantom \dagger}_{j}
-
\mu \sum_{i}
\hat{a}^{\dagger}_{i}  \hat{a}^{\phantom \dagger}_{i}
+
\frac{U}{2}  \sum_{i}
\hat{a}^{\dagger}_{i}  \hat{a}^{\dagger}_{i}  \hat{a}^{\phantom \dagger}_{i} \hat{a}^{\phantom \dagger}_{i}.
\end{equation}
\noindent
Here $\langle i,j \rangle$ denotes the sum over nearest neighbours, 
$t_{ij}$ are the hopping parameters,
and $\mu$ is the chemical potential.
Using our slave-boson operators we now rewrite Eq. (\ref{sep2}) into the form
\begin{eqnarray}\label{SHBhamiltonian}
\hat{H} = &-& 
\sum_{\langle i,j \rangle} 
\sum_{\alpha, \beta } \sqrt{\alpha +1} \sqrt{\beta +1}
(\hat{a}_{i}^{\alpha +1})^{\dagger} \hat{a}_{i}^{\alpha} 
t^{\phantom \dagger}_{ij} 
\hat{a}_{j}^{\beta + 1} (\hat{a}_{j}^{\beta})^{\dagger} 
- \mu 
\sum_{i} \sum_{\alpha } \alpha \hat{n}_{i}^{\alpha} 
\nonumber \\ 
&+&
\frac{U}{2} \sum_{i} \sum_{\alpha } \alpha (\alpha -1 ) \hat{n}_{i}^{\alpha},
\end{eqnarray} with the additional constraint given in Eq. (\ref{firstconstraint}).
We see that the quartic term in the original Bose-Hubbard Hamiltonian has been replaced by one that is quadratic in the 
slave-boson creation and annihilation operators, which is the most important 
motivation for the introduction of slave bosons. 

Now that we have introduced the slave-boson method and derived its 
representation of the Bose-Hubbard model, we want to turn the Hamiltonian 
into an action for the  imaginary time evolution. Using the standard 
recipe \cite{negele, stoofstf} we find
\begin{eqnarray}\label{sbfundamental}
S[(a^{\alpha}) ^{*},a^{\alpha},\lambda]
 &=& 
\int_{0}^{\hbar \beta} d \tau  \left\{ \sum_{i} \sum_{\alpha \beta}
(a^{\alpha}_{i})^{*} 
 M^{\alpha \beta} a^{\beta}_{i}
- i \sum_{i} \lambda_{i}(\tau) \left( \sum_{\alpha} n_{i}^{\alpha} -1 \right)  \right.
\nonumber \\ &-& \left.
\sum_{\langle i,j \rangle} 
\sum_{\alpha, \beta } \sqrt{\alpha +1} \sqrt{\beta +1}
(a_{i}^{\alpha +1})^{*} a_{i}^{\alpha} 
t_{ij} 
a_{j}^{\beta + 1} (a_{j}^{\beta})^{*} \right\}, 
\end{eqnarray} 
where $M$ is a diagonal matrix that
has as  the $\alpha^{th}$ diagonal entry
the term 
$ \hbar \partial / \partial \tau   -
	\alpha \mu +  \alpha (\alpha -1) U/2$, 
and $\beta= 1/k_{B}T$ is the inverse thermal energy.
The real valued constraint field $\lambda$ enters the action through,
\begin{equation}
\prod_{i} \delta(\sum_{\alpha} n_{i}^{\alpha} -1) = \int d[\lambda] e^{
\frac{i}{\hbar }  \int_{0}^{\hbar \beta} \sum_{i} 
\lambda_{i}(\tau) \left( \sum_{\alpha} n_{i}^{\alpha} -1 \right) 
d \tau}.
\end{equation}

Although we have simplified the interaction term, the hopping term has become more complicated. 
By performing a Hubbard-Stratonovich transformation on the above action we 
can, however, decouple the hopping term in a similar manner 
as in Ref. \cite{Oosten2001}. This introduces a field $\Phi$ into the action which, as we will 
see, may be identified with the superfluid order parameter. 
The Hubbard-Stratonovich 
transformation basically consists of adding a complete square to the action, i.e., adding
\begin{equation*}
\int_{0}^{\hbar \beta} d \tau \sum_{i,j}
\left( 
\Phi_{i}^{*} - \sum_{\alpha} \sqrt{\alpha+1} (a^{\alpha+1}_{i})^{*} 
a^{\alpha}_{i} 
\right)
t_{ij}
\left( 
\Phi_{j} - \sum_{\alpha} \sqrt{\alpha+1} a^{\alpha+1}_{i} 
(a^{\alpha }_{i})^{*} 
\right).
\end{equation*} Since a complete square can be added to the action without changing the physics we see that this
procedure allows us to decouple the hopping term.
We also perform
a Fourier transform on 
all fields by means of $a^{\alpha}_{i}(\tau) =  (1 / \sqrt{N_{s} \hbar \beta})
\sum_{{\bf k},n} a_{{\bf k},n}^{\alpha} e^{i ({\bf k} \cdot{\bf x}_{i} - \omega_{n} \tau)}$.
If we also carry out the  remaining integrals and sums 
we find
\begin{eqnarray}\label{3133}
&& S[\Phi^{*}, \Phi, (a^{\alpha})^{*},a^{\alpha},\lambda] = 
\sum_{{\bf k},n} \epsilon_{{\bf k}} | \Phi_{{\bf k},n} |^2 
- i  \frac{1}{\sqrt{N_{s} \hbar \beta}} \sum_{{\bf k},{\bf q}} \sum_{n,n'} \lambda_{{\bf q},n'} 
(a^{\alpha}_{{\bf k},n})^{*} a^{\alpha}_{{\bf k+q},n+n'} + 
i  N_{s} \hbar \beta \lambda \nonumber  \\
&&  
+ \sum_{{\bf k},n}
(a^{\alpha}_{{\bf k},n})^{*}
M^{\alpha \beta} (i \omega_{n} ) 
a^{\beta}_{{\bf k},n} 
-
\sum_{{\bf k},{\bf k'},n,n' }   
\frac{\epsilon_{{\bf k'}}}{\sqrt{N_{s} \hbar \beta}} 
 \left\{
\left(
\sum_{\alpha } \sqrt{\alpha +1}(a_{{\bf k+k'},n+n'}^{\alpha+1})^{*} a_{{\bf k},n}^{\alpha}
 \right) 
\Phi_{{\bf k'},n'} \right. \nonumber \\  &&+ \left.
\Phi^{*}_{{\bf k'},n'} 
\left(
\sum_{\alpha } \sqrt{\alpha +1} a_{{\bf k+k'},n+n'}^{\alpha+1} (a_{{\bf k},n}^{\alpha})^{*}
 \right) 
\right\}, \nonumber \\
\end{eqnarray} where the matrix $M(i \omega_{n}) $ is related to the matrix
$M$ in Eq. (\ref{sbfundamental}) through a Fourier transform.
Furthermore, $\lambda = ( \lambda_{{\bf 0},0}/ \sqrt{N_{s} \hbar \beta})$,
$\epsilon_{{\bf k}}  = 2 t \sum_{j=1}^{d} \cos{(k_{j} a)}$, where $a$ is the lattice 
constant of the square lattice with $N_{s}$ lattice sites.
For completeness we point out that the integration measure has become
\begin{equation}
\int d[(a^{\alpha})^{*} ] d[a^{\alpha}] = 
\int \prod_{{\bf k},n} d[(a^{\alpha}_{{\bf k},n})^{*} ] d[a^{\alpha}_{{\bf k},n}]  \frac{1}{\hbar \beta}.
\end{equation}

In principle Eq. (\ref{3133}) is still an exact rewriting of the Bose-Hubbard model.
As a first approximation we soften the constraint by replacing the 
general constraint field $\lambda_{i}(\tau)$ with a time and position independent field $\lambda$.
By neglecting the position dependence we enforce the constraint only on the sum of all lattice sites.
Doing this we are only left with the $\lambda_{{\bf 0},0}$ contribution in Eq. (\ref{3133}), 
which can then  be added to the matrix $M$. The 
path-integral  over the constraint field reduces to an ordinary integral. 
So we have,
\begin{subequations}
\begin{equation}
S[\Phi^{*}, \Phi, (a^{\alpha})^{*},a^{\alpha},\lambda] = S_{0} + S_{I}
\end{equation} where,
\begin{eqnarray}
S_{0} &=& 
i N_{s} \hbar \beta \lambda +
\sum_{\alpha, \beta} \sum_{{\bf k},n} \left\{ \epsilon_{{\bf k}} | \Phi_{{\bf k},n} |^2 
+ 
(a^{\alpha}_{{\bf k},n})^{*}  \ 
M^{\alpha \beta}(i \omega_{n})  \ 
a^{\beta}_{{\bf k},n} 
\right\} \equiv 
S^{SB}_{0} + \sum_{{\bf k},n} \epsilon_{{\bf k}} | \Phi_{{\bf k},n} |^2 ,
\end{eqnarray} The matrix
$M^{\alpha \beta }(i \omega_{n}) = \delta_{\alpha \beta} (- i \hbar \omega_{n} - i 
\lambda  - \alpha \mu + \alpha (\alpha - 1) U/2)$, and
\begin{eqnarray}
S_{I} = &&-
\sum_{{\bf k},{\bf k'},n,n' }   
\frac{\epsilon_{{\bf k'}}}{\sqrt{N_{s} \hbar \beta}} 
 \left\{
\left(
\sum_{\alpha } \sqrt{\alpha +1}(a_{{\bf k+k'},n+n'}^{\alpha+1})^{*} a_{{\bf k},n}^{\alpha}
 \right) 
\Phi_{{\bf k'},n'} \right. \nonumber \\  &&+ \left.
\Phi^{*}_{{\bf k'},n'} 
\left(
\sum_{\alpha } \sqrt{\alpha +1} a_{{\bf k+k'},n+n'}^{\alpha+1} (a_{{\bf k},n}^{\alpha})^{*}
 \right) 
\right\}. \nonumber \\
\end{eqnarray}

\end{subequations}

The crucial idea of Landau theory is that near a critical point the  quantity of most interest is the order parameter.
In our theory the superfluid field $\Phi$ plays the role of the order 
parameter. Only $\Phi_{{\bf 0},0}$ can have a nonvanishing expectation value in our case and, therefore,
we can write the ground-state energy as an expansion in powers of $\Phi_{{\bf 0},0}$,
\begin{equation}\label{gsenergy}
E_{g} (\Phi_{{\bf 0},0} ) = a_{0} (\alpha, U ,\mu) + a_{2} (\alpha , U ,\mu) | \Phi_{{\bf 0},0} |^2 +
\mathcal{O}( | \Phi_{{\bf 0},0} |^4 ),
\end{equation}

\noindent
and minimize it as a function of the superfluid order parameter $\Phi_{{\bf 0},0}$. We  thus find that 
$\langle \Phi_{{\bf 0},0} \rangle = 0$
when $a_{2} (\alpha , U, \mu)  > 0$ and that $\langle \Phi_{{\bf 0},0} \rangle \neq 0$ when $a_{2} (\alpha, U , \mu) < 0$. 
This means that 
$a_{2} (\alpha , U, \mu) = 0 $ signals the boundary between the superfluid and the insulator phases at zero 
temperature  and the boundary between the superfluid and the normal phases at nonzero temperature.
Therefore we are going to calculate the effective action of our theory up to second order in $\Phi$.
The zeroth-order term in the expansion of the action in powers of the order parameter gives us the zeroth-order 
contribution $\Omega_{0}$ to the thermodynamic potential $\Omega$.  We have,
\begin{eqnarray}
e^{-\beta \Omega_{0} } \equiv 
\int \prod_{\alpha} \left( \prod_{{\bf k},n} 
	d[(a^{\alpha}_{{\bf k},n})^{*} ] d[a^{\alpha}_{{\bf k},n}]  \frac{1}{\hbar \beta} \right) 
	e^{-S^{SB}_{0} / \hbar} .
\end{eqnarray} From this it follows that, 
\begin{equation}
-\beta \Omega_{0} = -i N_{s} \beta \lambda + N_{s} \sum_{\alpha}  
\log{\left( 1- e^{-\beta
M^{\alpha \alpha}(0)
}\right)},
\end{equation} and  
$M^{\alpha \alpha}(0) =  (- i \lambda - \alpha \mu +  \alpha (\alpha -1)U/2 )$.
Next we must calculate $\langle S_{I}^{2} \rangle$
where $\langle \cdots \rangle$ denotes averaging with respect to $S_{0}$, i.e.,
\begin{equation}
\langle A \rangle  = \frac{1}{e^{-\beta \Omega_{0}}}
\int \prod_{\alpha} \left( \prod_{{\bf k},n} 
	d[(a^{\alpha}_{{\bf k},n})^{*} ] d[a^{\alpha}_{{\bf k},n}]  \frac{1}{\hbar \beta} \right) 
	A[(a^{\alpha})^{*},a^{\alpha}] e^{-S_{0}^{SB} / \hbar} .
\end{equation} Once we have this contribution, we 
automatically also find the dispersion relations for the quasiparticles in our 
system as we will see shortly. For small $\Phi$ we are allowed to expand the exponent
in the integrand of the functional integral for the partition function as
\begin{equation}\label{effact}
e^{-S/\hbar} = e^{-(S_{0} + S_{I})/\hbar} \approx e^{-S_{0} / \hbar} (1 -
 S_{I}/\hbar + \frac{1}{2}
(S_{I} /\hbar)^2).
\end{equation} 
It can be shown that the expectation value of $S_{I}$ vanishes. 
The second order contribution is found to be,
\begin{equation}
\langle S_{I}^2 \rangle =
2
\sum_{{\bf k},{\bf k'},n,n'}
\epsilon_{{\bf k}}^2 
 \frac{|\Phi_{{\bf k}}|^2}{N_{s} \hbar \beta}
 \sum_{\alpha } (\alpha +1) 
\langle
(a^{\alpha + 1}_{{\bf k + k'},n + n'} )^{*}
a^{\alpha + 1}_{{\bf k + k'},n + n'} 
\rangle
\langle
(a^{\alpha }_{{\bf k}, n } )^{*}
a^{\alpha }_{{\bf k},n } 
\rangle.
\end{equation}

One of the sums over the Matsubara frequencies $\omega_{n}$ 
can be performed and the sum over ${\bf k'}$ produces 
an overal  factor $N_{s}$.
We thus find 
\begin{equation}
 \langle S_{I}^2 \rangle =
\sum_{{\bf k},n }
\epsilon_{{\bf k}}^2 
 \frac{|\Phi_{{\bf k}}|^2}{\hbar \beta}
 \sum_{\alpha } (\alpha +1) 
\frac{n^{\alpha} - n^{\alpha + 1}}{- i \hbar \omega_{n} - \mu + \alpha U},
\end{equation} where we have defined the occupation numbers  $n^{\alpha} \equiv \langle (a^{\alpha  }_{i})^{*}  
a^{\alpha}_{i} \rangle$
that equal
 \begin{equation}\label{occnodef}
n^{\alpha} =
 \frac{1}{\exp{\left\{ \beta \left( -i \lambda - \alpha \mu + \frac{1}{2} \alpha (\alpha -1) U 
\right) \right\}} -1}. 
\end{equation} 

\noindent
Having performed the integrals over the slave-boson fields to second order, we can 
exponentiate the result to obtain the effective action for the order parameter
\begin{eqnarray}\label{effaction}
S^{\rm eff}[\Phi^{*},\Phi] &=& 
\left( \hbar \beta \Omega_{0} - \hbar  \sum_{{\bf k},n}  \Phi^{*}_{{\bf k},n} G^{-1}({\bf k},i \omega_{n}) 
\Phi_{{\bf k},n} 
\right),
\end{eqnarray}where we have defined the Green's function
\begin{equation}\label{groenefunctie}
- \hbar G^{-1} ({\bf k}, i \omega_{n})= 
\left(
\epsilon_{{\bf k}} - \epsilon_{{\bf k}}^2 
 \sum_{\alpha } (\alpha +1) 
\frac{n^{\alpha} - n^{\alpha + 1}}{- i \hbar \omega_{n} - \mu + \alpha U}
\right).
\end{equation}

This result is one of the key results of this paper, which is correct in the limit of small $\Phi_{{\bf k},n}$.
If we want to make the connection with the Landau theory again, we can identify the 
$a_{2} (\alpha, U, \mu)$ in Eq. (\ref{gsenergy}) with $G^{-1}({\bf 0},0)/\beta$. In Sec. \ref{meanfieldresults}
we analyse this further.

\subsection{Mott insulator}

In the Mott insulator where $n_{0} \equiv | \langle \Phi_{{\bf 0},0} \rangle |^2 = 0$, the thermodynamic potential is now 
easily calculated by integrating out the superfluid field. In detail
\begin{eqnarray}
Z \equiv e^{-\beta \Omega} &= &\int d \lambda d[\Phi^{*}] d[\Phi] e^{-S^{\rm eff}/\hbar} \nonumber \\
&=& 
\int d \lambda  \exp{\left\{ - \beta \Omega_{0} 
 - \sum_{{\bf k},n} \log{ \left[ \beta \left(   
\epsilon_{{\bf k}} - \epsilon_{{\bf k}}^2 
 \sum_{\alpha } (\alpha +1) 
\frac{n^{\alpha} - n^{\alpha + 1}}{- i \hbar \omega_{n} - \mu + \alpha U}
\right) \right] }
\right\}}.
\end{eqnarray}
\noindent
At this point we perform a saddle point approximation for the constraint field $\lambda$. 
This implies that we only take into account that value of  
$\lambda$ that maximizes the canonical partition function.
If we now  thus minimize the free energy with respect to the chemical potential and the constraint field, 
we get two equations that need to be solved.
The first is $\partial \Omega / \partial \lambda = 0$ and reads,
\begin{subequations}\label{conseqn}
\begin{equation}
N_{s} \left(  1 -   \sum_{\alpha} n^{\alpha} \right) - \frac{i}{\beta}  \sum_{{\bf k},n} G({\bf k}, i \omega_{n}) \frac{\partial  
G^{-1}({\bf k},i \omega_{n})}{\partial \lambda} =0 .
\end{equation} In a mean-field approximation the last term is neglected, and this equation tells 
us that the sum of the average slave-boson occupation numbers 
must be equal to one. This reflects the constraint of one slave boson per site.
The second equation follows from $- \partial \Omega/ \partial \mu = N$ and gives
\begin{equation}
  N_{s} \sum_{\alpha} \alpha n^{\alpha}  + \frac{1}{\beta} \sum_{{\bf k},n} G({\bf k},i \omega_{n}) 
\frac{\partial G^{-1}({\bf k},i \omega_{n})}{\partial \mu	} = N.
\end{equation}
\end{subequations} This equation shows how the particle  density can be seen as the sum of terms $\alpha 
n^{\alpha}$ and a correction coming from the propagator of the superfluid order parameter.
 The latter is again neglected in the mean-field approximation.

\subsection{Superfluid phase}

In the superfluid phase the order parameter $| \Phi_{{\bf 0},0} |^2$ has a nonzero expectation value.
We find this expectation value by calculating the minimum of the classical part of the action, i.e.,
$ - \hbar G^{-1}({\bf 0},0)  | \Phi_{{\bf 0},0} |^2 + a_{4} | \Phi_{{\bf 0},0} |^4 $. This minimum 
becomes nonzero when $-\hbar G^{-1}({\bf 0},0)$ becomes negative, and is then equal to 
\begin{equation}
| \langle \Phi_{{\bf 0},0} \rangle |^2 = \frac{ \hbar G^{-1}({\bf 0},0)}{2 a_{4}} \equiv n_{0}
\end{equation}

In appendix \ref{higherorder} we calculate the coefficient $a_{4}$ of the fourth order term 
$| \Phi_{{\bf 0},0}|^4$. We approximate  the prefactor to the fourth order term, which in general depends 
on momenta and Matsubara frequencies,  with the zero-momentum and zero-frequency 
value of $a_{4}$ so that
the approximate action to fourth order becomes,
\begin{equation}\label{fourthorderaction}
S = \hbar \beta \Omega_{0} -\hbar \sum_{{\bf k},n}   \Phi_{{\bf k},n}^{*} G^{-1}({\bf k},i \omega_{n}) 
\Phi_{{\bf k},n} 
+ a_{4} \sum_{{\bf k,k',k''}} \sum_{n,n',n''}  \Phi_{{\bf k},n}^{*} \Phi_{{\bf k'},n'}^{*} \Phi_{{\bf k''},n''} \Phi_{{\bf k + k' - 
k''},n+n' -n''}  
\end{equation}
We now write the order parameter as the sum of its expectation value plus fluctuations, i.e.,
\mbox{$\Phi_{{\bf 0},0} 
\rightarrow  \sqrt{n_{0}} \cdot \sqrt{N_{s} \hbar \beta} + \Phi_{{\bf  0},0}$}  and  
a similar expression  for  $\Phi^{*}_{{\bf 0},0}$.
If we put this into the action and only keep the terms up to second order, the contribution of the
fourth-order term is given by
\begin{equation*}
 a_{4} n_{0}  \sum_{{\bf k},n} \left( \Phi_{{\bf k},n} \Phi_{{\bf -k},-n} + 4 \Phi^{*}_{{\bf k},n} \Phi^{\phantom 
*}_{{\bf k},n} + \Phi^{*}_{{\bf k},n} \Phi^{*}_{{\bf -k},-n} \right).
\end{equation*}There is also a contribution $-\hbar G^{-1}({\bf 0},0) n_{0}$ from the second-order term.
To summarize, in the superfluid phase we can write 
the action Eq. (\ref{fourthorderaction}) to second order as
\begin{eqnarray}
&& S^{\rm SF} = 
\hbar \beta \Omega_{0} - \hbar G^{-1}({\bf 0},0) n_{0} -\frac{\hbar}{2}
\sum_{{\bf k},n} 
\left(\begin{matrix}
\Phi_{{\bf k},n}^{*} &  \Phi_{{\bf -k},-n}
\end{matrix}\right)
\bf{G^{-1}}({\bf k},i \omega_{n})
\left(\begin{matrix}
\Phi_{{\bf k},n} \\ \Phi_{{\bf -k},-n}^{*}
\end{matrix}\right) \nonumber \\
&& - {\bf G^{-1}}({\bf k}, i \omega_{n} ) = 
\left(\begin{matrix}
-   G^{-1}({\bf k},i \omega_{n}) + 4 \hbar  a_{4} n_{0} &  2 \hbar a_{4} n_{0} \\
2 \hbar a_{4} n_{0} & -  G^{-1}(-{\bf k},-i \omega_{n}) + 4 \hbar a_{4} n_{0}  
\end{matrix}\right).
\end{eqnarray}

\noindent
Integrating out the field $\Phi_{{\bf k},n}$ we find the Bogoliubov expression for the thermodynamic potential in the 
superfluid phase,
\begin{eqnarray}
Z \equiv e^{-\beta \Omega} &= &\int d \lambda d[\Phi^{*}] d[\Phi] e^{-S^{\rm SF}/\hbar} \nonumber \\ 
&=& \int  d \lambda \exp{\left\{ -\beta \Omega_{0} +  n_{0} G^{-1}({\bf 0},0) 
-\rm{Tr}{\left[ \log{(- \hbar \beta \bf{G^{-1}})} \right]} 
\right\}}
\end{eqnarray}

\section{Mean-field Theory}\label{meanfieldresults}

In this section, we apply the theory we have developed in the previous section.
First, using the Landau procedure, we reproduce the mean-field zero-temperature phase diagram.
We then study the phase diagram at nonzero temperatures. To do so  
we calculate the compressibillity of our system as a function of temperature,  showing how 
for fixed on-site repulsion $U$ the Mott insulating region gets smaller. 
By also looking at the condensate density as a function of temperature, we get a quantitative picture of 
what happens 
at fixed on-site repulsion $U$. The nice feature is that all our expressions are analytic.
Next, we consider our system at zero temperature again and we study at the mean-field level the behaviour of the 
compressibillity  as we go from the superfluid phase to the Mott insulating phase. 
What we find is consistent with the general idea that  the quantum phase transition between the Mott insulator and the 
superfluid phases belongs to different universality classes depending on how you walk through the phase diagram
 (cf. Ref. \cite{Sachdev}).
We then obtain an analytic expression for the critical temperature of the superfluid-normal phase transition in the 
approximation of three slave bosons, i.e., up to doubly-occupied sites. Numerically we extend this study to include 
a fourth  slave boson and find only slight changes to $T_{c}$.
From the propagator of the superfluid field we  extract the dispersion relations of the quasiparticle-quasihole pairs
and their temperature dependence.

\subsection{Zero-temperature phase-diagram}\label{meanfieldresultsA}

\begin{figure}
\includegraphics[width=10cm]{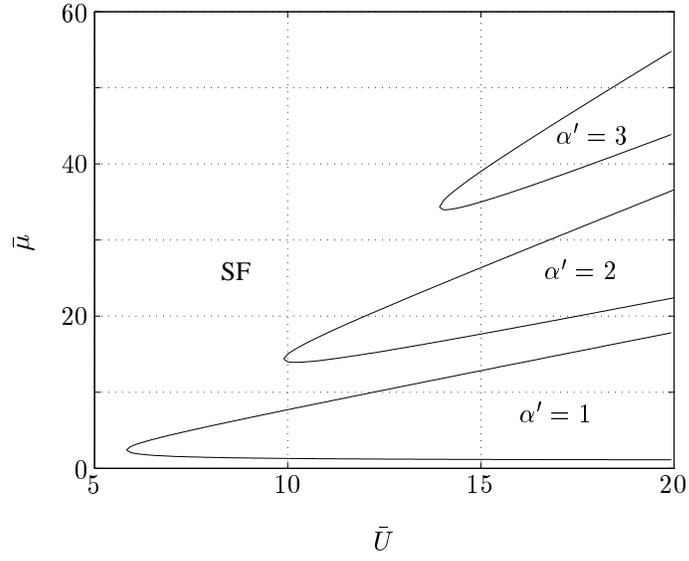}
\caption{Phase diagram of the Bose-Hubbard Hamiltonian as obtained from 
the mean-field zero-temperature limit in the slave-boson formalism. It 
shows the superfluid (SF) phase and the Mott insulator regions with 
different integer filling factors here denoted by $\alpha'$.
The vertical axis shows the dimensionless chemical potential $\bar{\mu} = \mu/zt$
and the horizontal axis shows the dimensionless interaction strength $\bar{U} = U/zt$.}\label{fig2}
\end{figure}

From the zeros of  $G^{-1}({\bf 0},0)$ in  Eq. (\ref{groenefunctie}),  we obtain the mean-field phase diagram
in the $(\mu, U)$~plane.
For a Mott insulating state with integer filling factor $\alpha'$ we have $n^{\alpha} = \delta_{\alpha,\alpha'}$.
When this is substituted into the equation $G^{-1}({\bf 0},0)= 0$  we can find the $U(\mu)$ 
curve that solves that equation
and thus  determines the size of this Mott insulating state. 
For given filling factor $\alpha'$ we also define $U_{c}$ as the minimal $U$ that solves the equation. 
Within the Mott insulating phase we have a zero compressibility
$\kappa \equiv \partial n / \partial \mu $, where $n=n(\mu,U)$ is the total density as determined from
the thermodynamic potential.
Straightforward calculation gives that we are in a Mott insulating phase  whenever $\bar{\mu}$ lies between 
$\bar{\mu}_{-}^{\alpha'}$ and $\bar{\mu}_{+}^{\alpha'}$ 
where,
\begin{equation}\label{critmu}
\bar{\mu}_{\pm}^{\alpha'} = \frac{1}{2} \left( \bar{U} (2 \alpha' -1) -1  \right) \pm
\frac{1}{2} \sqrt{\bar{U}^2 - 2 \bar{U} (2 \alpha' + 1) + 1}.
\end{equation}Here we have introduced the dimensionless chemical potential $\bar{\mu} \equiv \mu / z t $  and on-site repulsion
strength $\bar{U} \equiv U/zt$.
When $\bar{\mu}$ does not lie between any $\bar{\mu}^{\alpha'}_{-}$ and $\bar{\mu}_{+}^{\alpha'}$ the 
`superfluid' density  $|\langle \Phi_{{\bf 0},0} \rangle |^2  $ will no 
longer be  zero and the Mott insulating phase has disappeared. We have drawn the zero temperature phase 
diagram in Fig.
\ref{fig2}. Our slave-boson approach reproduces here the results of previous mean-field studies 
\cite{Fisher1989, Sheshadri1993, Oosten2001}.
For nonzero temperatures the equation $G^{-1}({\bf 0},0) = 0$  no longer describes a quantum phase transition between a 
superfluid and a 
Mott insulator but it describes a thermal phase  transition between a superfluid and a normal phase. We will look into this in 
more detail in Sec. 
\ref{sf-n}.

\subsection{Compressibillity}\label{sec3b}

\begin{figure}
\begin{center}
\subfigure[]{\includegraphics[width=8cm]{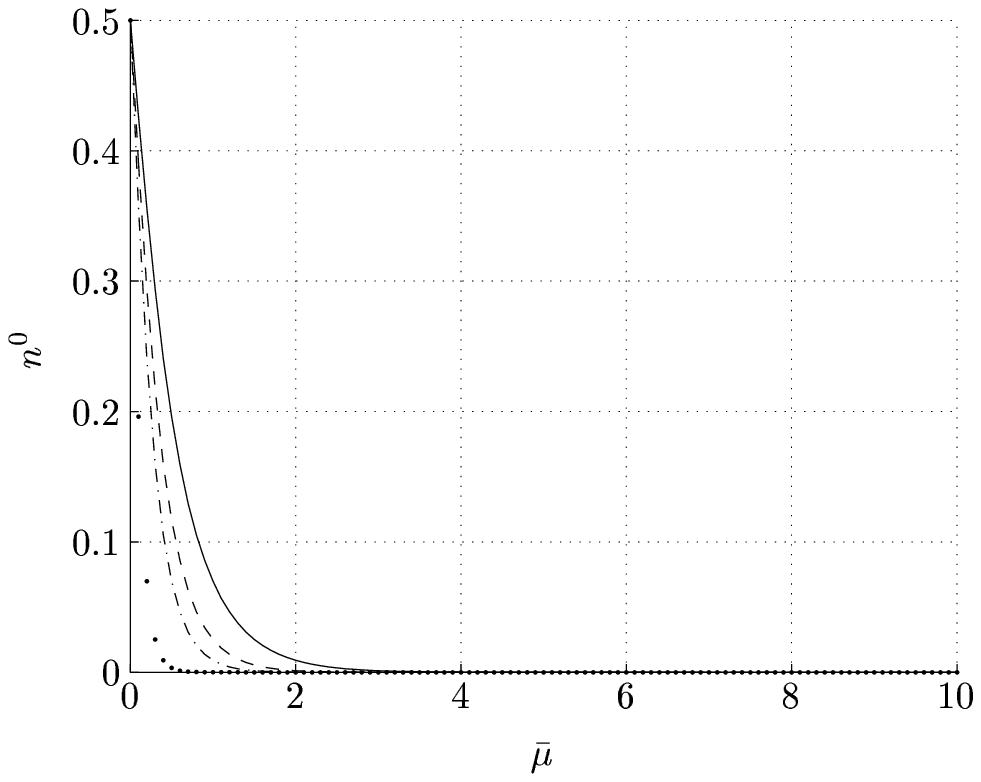}}
\goodgap
\subfigure[]{\includegraphics[width=8cm]{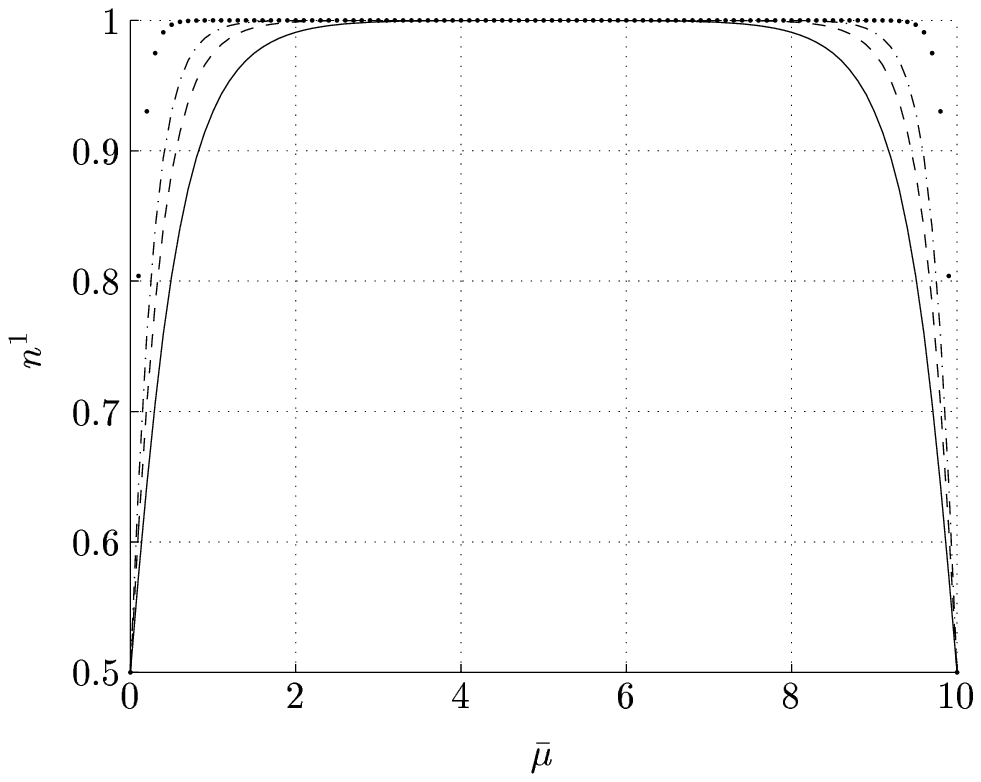}}
\goodgap \\
\subfigure[]{\includegraphics[width=8cm]{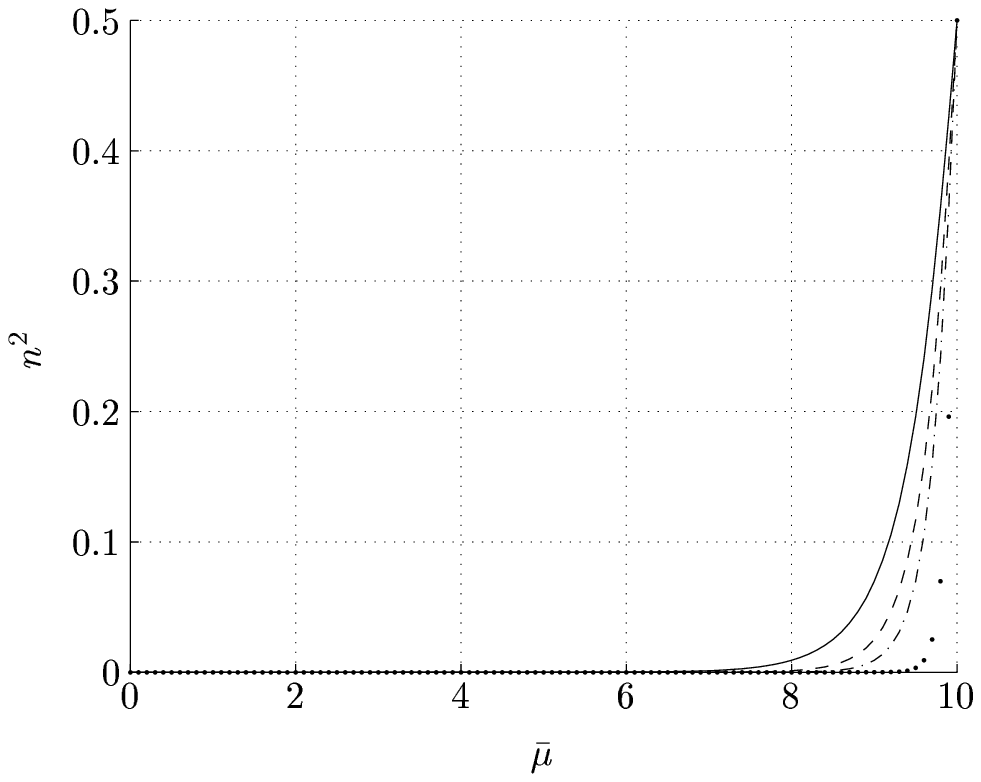}} 
\goodgap
\subfigure[]{\includegraphics[width=8cm]{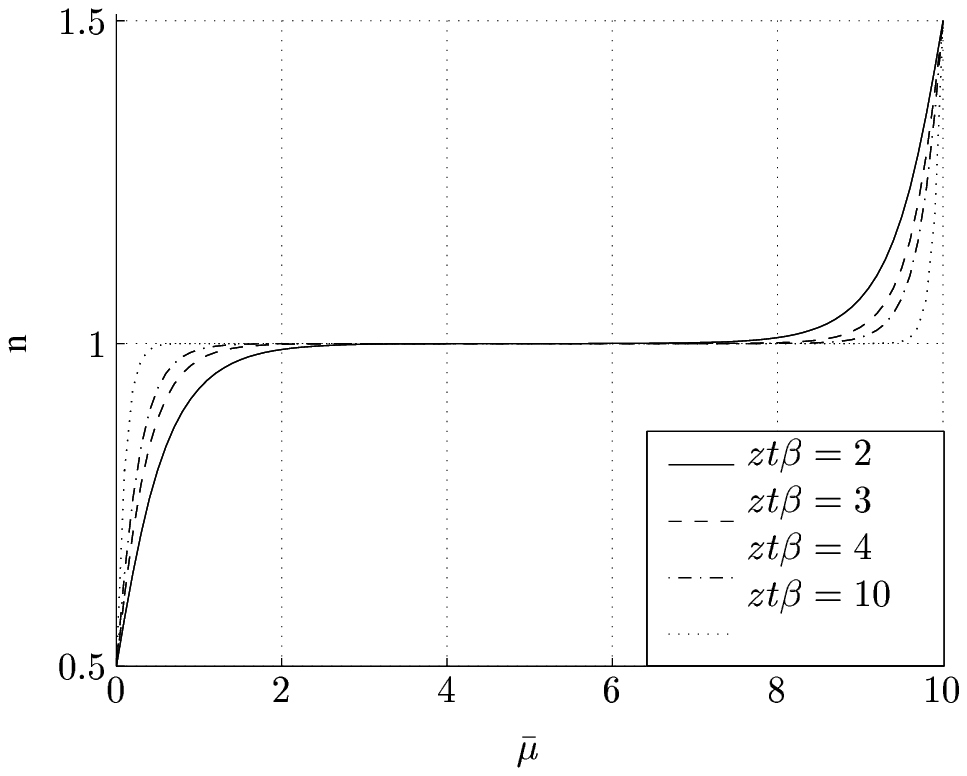}}
\goodgap
\caption{Numerical solution of the slave-boson occupation numbers $n^{0},n^{1}$ and $n^{2}$ 
is shown in Figs. (a)-(c)
as a function of $\bar{\mu}$ for various temperatures and
 for fixed $U/zt =10$. Figure (d) shows the 
total density $n$. As a function of temperature the compressibillity increases.
} \label{diffoccno}
\end{center}
\end{figure}

To see what happens to the Mott insulator as we move away from zero 
temperature we must look at the compressibillity as a function of 
temperature. Numerically we have solved Eq.
(\ref{conseqn}), which gives us the occupation numbers of the slave bosons as 
depicted in Fig. \ref{diffoccno}. With that we can determine the total 
density in the phase where the order parameter is zero.
It is clear that within a mean-field approximation the compressibillity at 
zero  temperature is exactly zero. In Fig. \ref{diffoccno} we have 
plotted the total density as a function of temperature. As the temperature is 
raised we see that the compressibillity, which is the slope of the curve,  becomes nonzero.
This shows that there is no longer a Mott insulator present.
We also see that even though the slope is no longer zero it is very 
small indeed for low enough temperatures. We determine the crossover 
line by requiring that  $\Delta (T)/k_{B} T$ is of order one, 
where $\Delta (T)$ is defined  as  the difference of the
quasiparticle and quasihole dispersions at ${\bf k} = 0$.

\subsection{Superfluid density}\label{sfdens}

\begin{figure}
\includegraphics[width=10cm]{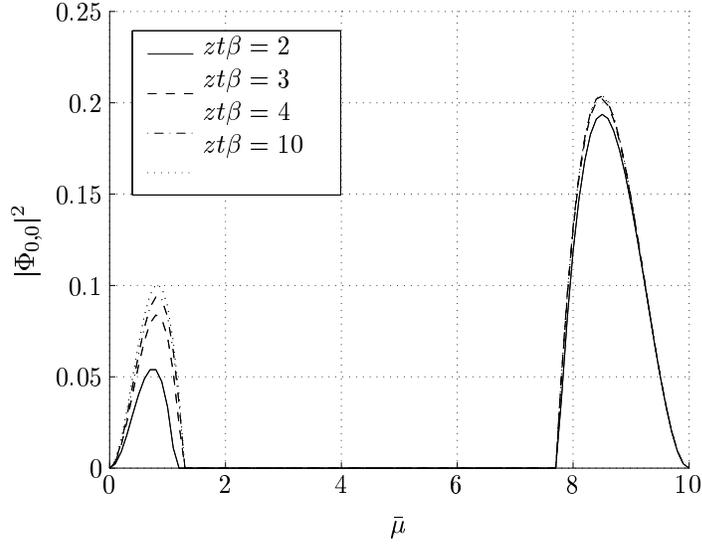}
\caption{Superfluid density $| \Phi_{{\bf 0},0} |^2$ as a function of $\bar{\mu}$  for various 
temperatures and for $U/zt=10$. The superfluid density as well as the region of superfluid phase 
diminish as a function of increasing temperature. The vanishing of $| \Phi_{{\bf 0},0} |^2 $ at $\bar{\mu} = 0$ and 
$\bar{\mu} = 10$ is an artefact of our approximation (see text).}
\label{sfdens2}
\end{figure}

In a mean-field approximation the superfluid density is 
extracted from the action by finding the $|\langle \Phi_{{\bf 0},0} \rangle |^2$ that minimizes the fourth-order 
action in
Eq. (\ref{fourthorderaction}),
\begin{equation}
|\langle \Phi_{{\bf 0},0} \rangle |^2 = \frac{\hbar G^{-1}({\bf 0},0)}{2 a_{4}} ,
\end{equation} whenever $\mu$ is not between $\mu_{-}^{\alpha'}$ and $\mu_{+}^{\alpha'}$, and zero 
otherwise. We have plotted this expectation value in Fig. \ref{sfdens2} for $\alpha' = 1$.
In this figure we see how the superfluid density grows as a function of $\mu$ moving away from
the Mott insulator phase. Our expansion of the Landau free energy
 is only valid around the edge of the Mott lobes and therefore 
breaks down when we go too far away from the Mott insulator. This can be seen in the figure as the decrease of the superfluid 
density when $\mu$ approaches $0$ and/or $U$.
It can also be seen from the propagator of the superfluid 
field, which has poles when $\mu = \alpha U$.
For $U$ not too far away from the insulating phase the figure 
quantitatively agrees with the ones calculated  by other authors
\cite{Oosten2001}.

\subsection{Bogoliubov dispersion relation}

We now demonstrate that the dispersion $\hbar \omega_{{\bf k}}$ is linear 
in ${\bf k}$ in the superfluid phase and that the spectrum is gapless.
In the superfluid phase we can expand around the expectation value $n_{0} = \hbar G^{-1}({\bf 0},0)/ 2a_4 $  
of the order parameter. Up to quadratic-order this gives,
\begin{equation}
S = \hbar \beta \Omega_{0} - \hbar  \sum_{{\bf k},n} \Phi^{*}_{{\bf k},n} G^{-1}({\bf k}, i \omega_{n}) 
\Phi_{{\bf k},n} + a_{4} n_{0} 
\sum_{{\bf 
k},n} \left(
\Phi_{{\bf k},n} \Phi_{{\bf -k},-n} + 4 \Phi^{*}_{{\bf k},n} 
\Phi^{\phantom *}_{{\bf k},n} + \Phi^{*}_{{\bf k},n} \Phi^{*}_{{\bf -k},-n} \right).
\end{equation} From this we find the dispersion-relation $\hbar \omega_{{\bf k}}$ in the superfluid in the usual way.
We perform an analytic continuation $G^{-1}({\bf k},i \omega_{n}) \rightarrow G^{-1}({\bf k}, \omega_{k})$ and find
\begin{equation}
\hbar \omega_{{\bf k}} = \hbar 
\sqrt{\left(  G^{-1}({\bf k},  \omega_{\bf k}) /2  - G^{-1}({\bf 0},0) \right)^2 - \left( G^{-1}({\bf 0},0) / 2\right)^2}.
\end{equation} Note that $({\bf k},\omega_{\bf k}) = ({\bf 0},0)$ is a solution. Expanding around this solution in  
${\bf k}$  now gives,
\begin{equation}
\hbar \omega_{{\bf k}} = a \frac{\hbar G^{-1}({\bf 0},0)}{ \sqrt{2}  }  |{\bf k}|, 
\end{equation} where $a$ is again the lattice constant.
\subsection{Near the edges of the Mott lobe}

If we substitute the vacuum expectation value of the order parameter  back 
into our effective action, we see that the zeroth-order contribution to the 
thermodynamic potential in the superfluid phase in mean-field approximation is given by,
\begin{equation}
\hbar \beta \Omega =  \hbar \beta \Omega_{0} - \frac{\left(\hbar G^{-1}({\bf 0},0)\right)^2}{2 a_{4}}.
\end{equation} From this the particle density can be obtained by making use of the
thermodynamic identity $N = - \partial \Omega /\partial \mu$.
We can calculate this at $T=0$ and take the limit $\mu \rightarrow
\mu_{\pm}^{\alpha'}$ to show that the derivative of the density with respect to
$\mu$, i.e, $\partial n /\partial \mu$ shows a kink for all $U \neq U_c$.
This means that only if we walk through the tip of the Mott lobes
there is not a kink in the compressibility.
In fact it's not hard to see why this is true. At zero temperature the roots of 
$-\hbar G^{-1}({\bf 0},0)$ are by definition $\mu_{\pm}^{\alpha'}$. This means that we can
write $- \hbar G^{-1}({\bf 0},0) = C (\mu - \mu_{-}^{\alpha'}) (\mu - \mu_{+}^{\alpha'})$. The 
proportionality
constant can be shown to be equal to $C = \epsilon_{\bf 0} /((\alpha' U - \mu) ((\alpha' -1) U - 
\mu ))$.
This then shows that the thermodynamic potential is,
\begin{equation}
\hbar \beta \Omega = \hbar \beta \Omega_{0} +  \frac{C^2}{4} \frac{\left( \mu - \mu_{-}^{\alpha'} \right)^2
\left( \mu - \mu_{+}^{\alpha'} \right)^2 }{a_{4}}.
\end{equation} Remembering that the density is the derivative of the thermodynamic
potential we see that the second derivative of the thermodynamic potential
with respect to $\mu$ can show a nonzero value upon approaching the Mott
lobe. Since in the Mott isolator the density is constant and equal to $\alpha'$
we have shown the existence of a kink in the slope of the density for all
paths not going through the tip of the Mott lobe.
This causes the difference in the universality class of the quantum phase transition.

\subsection{The superfluid-normal phase transition}\label{sf-n}

In this subsection, we show that it is possible to obtain an analytical
expression for the critical temperature $T_{\rm c}$ of the transition
between superfluid and normal phases as a function of $U$, 
for values of $U$ below the critical $U$ of the zero-temperature  superfluid-Mott insulator transition.
The analytical result is obtained if we include occupations up to two
per site, i.e.,  three  slave bosons or occupation numbers $n^0, n^1, n^2$.
Along similar lines $T_{\rm c}$ can be found numerically if more slave 
bosons are included. We have carried out this procedure for the case of adding 
a fourth boson (triple occupancy) and find only modest quantitative
changes.

If we restrict the system to occupancies 0, 1 and 2, and fix the total
density $n \equiv N/N_{s}$ at 1, the occupation numbers $n^0, n^1$ and $n^2$ should obey
the following relations if we neglect fluctuation corrections (cf. Eq. (\ref{conseqn})):
\begin{eqnarray}
&& n^0 + n^1 + n^2 = 1 \label{eq:n012}, \\
\text{and} \nonumber \\
&& n^1 + 2 n^2 = 1  \label{eq:n122}.
\end{eqnarray} The $n^\alpha$ are furthermore given by Eq. (\ref{occnodef}), enabling us to 
eliminate $\lambda$ and express $n^0$ and $n^2$ in terms of $n^1$. We obtain
\begin{eqnarray}
 n^0 & = & \frac{n^1}{\left(n^1 + 1\right)\exp(\beta\mu) - n^1} \label{eq:n0},  \\
\text{and} \nonumber \\
 n^2 & = & \frac{n^1}{\left(n^1 + 1\right)\exp(\beta(U-\mu)) - n^1} \label{eq:n2}.
\end{eqnarray} The constraints in Eqs. (\ref{eq:n012}) and (\ref{eq:n122}) immediately lead to $n^0 = n^2$, 
so that,
according to Eqs. (\ref{eq:n0}) and (\ref{eq:n2}), we must have $\mu = U/2$. We notice that
at this level of approximation, we obtain a slight discrepancy with the
result from Sec. \ref{meanfieldresultsA} that at zero temperature  the critical value of 
$\bar{U}$ of  the superfluid-Mott  insulator
transition, which is the limiting $\bar{U}$ for the superfluid-normal transition that is
addressed here, is according to Eq. (\ref{critmu}) with 
$\alpha' = 1$ determined by  $\bar{\mu} = (\bar{U}-1)/2$  \cite{remedy}.

As argued above the criticality condition for the superfluid-normal transition is obtained
by putting $G^{-1}({\bf 0}, 0)= 0$. Restricting the
sum in the right-hand side of  Eq. (\ref{groenefunctie})  to $\alpha=0$ and $\alpha=1$, we obtain 
\cite{n3mn2}
\begin{equation}
1 = \frac{2}{\bar{\mu} - \bar{U}}\left( n^2 - n^1 \right) + 
    \frac{1}{\bar{\mu}} \left( n^1 - n^0 \right) .
\label{eq:critc}
\end{equation} Since the relation between
$\mu$ and $U$ is fixed by Eqs. (\ref{eq:n012}) and (\ref{eq:n122}), and 
$n^0$ and $n^2$ can be  expressed in $n^1$  as $n^0 = n^2 = (1-n^1)/2$, 
the criticality condition 
Eq. (\ref{eq:critc}) results in a remarkably simple relation between 
$n^1$ and $\bar{U}$ at $T_{\rm c}$, namely  $n^1 = (\bar{U}+3)/9$. Using this in Eq. (\ref{eq:n0})
leads to the following analytic formula for $\bar{T}_{\rm c} \equiv T_{c} / zt $ for the superfluid-normal transition:
\begin{equation}
k_{\rm B} \bar{T}_{\rm c} = 
\frac{\bar{U}}{2} \, \log^{-1} \left[\frac{(\bar{U}-24)(\bar{U}+3)}{(\bar{U}-6)(\bar{U}+12)} \right] ~.
\label{eq:Tc3b1}
\end{equation}
It is straightforward to generalize this procedure to arbitrary integer
density $\alpha'$ while allowing occupation numbers $n^{\alpha'-1}, n^{\alpha'},
n^{\alpha'+1}$ only. The result is
\begin{equation}
k_{\rm B} \bar{T}_{\rm c}^{\alpha'} =
\frac{\bar{U}}{2} \, 
\log^{-1} \left[\frac{(\bar{U}-8 (2\alpha' + 1))(\bar{U}+(2\alpha' + 1))}{(\bar{U}-2(2\alpha' + 1))(\bar{U}+4(2\alpha' + 
1))} \right] .
\label{eq:Tc3ba}
\end{equation}

\begin{figure}[htb!]
\includegraphics[width=10cm]{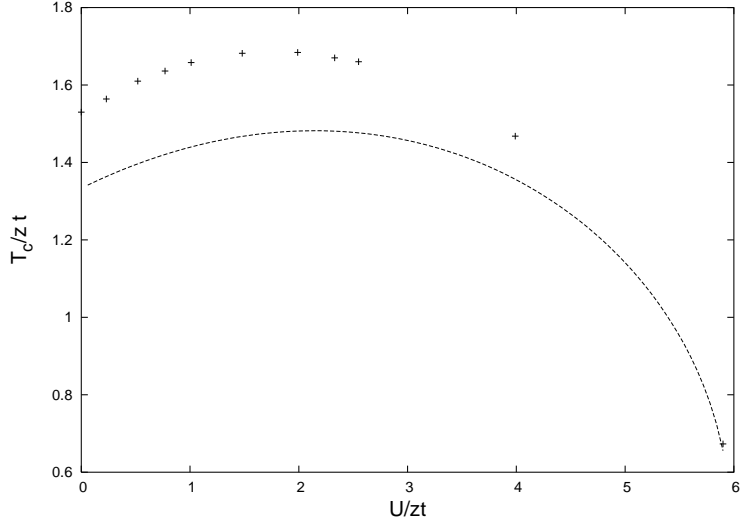}
\caption{ Critical temperature $T_{c}$  of the superfluid-normal phase transition
as a function of the interaction strength $U/zt$.
The solid line is an analytic expression obtained in  the approximation where we only 
take into account three slave bosons. The plusses correspond to a numerical solution 
for the case of four slave bosons.}\label{4bosons.eps}
\end{figure}

The critical temperature 
$T_{\rm c}$ for integer filling factor  $n \equiv N/N_{s} =1$, i.e., Eq. (\ref{eq:Tc3b1}), is plotted in Fig. \ref{4bosons.eps}.
The overall qualitative behavior is as one would expect (cf. Fig. \ref{qualphasediag}).
A few finer details appear to be less satisfactory. For instance, $T_{\rm c}$
vanishes for $\bar{U}=6$, whereas we would expect this to coincide with
the mean-field result for $\bar{U}_{\rm c}$ for the 
superfluid-Mott insulator transition for the
first Mott lobe, i.e., $\bar{U}_{\rm c} = 5.83$ obtained from  Eq. (\ref{critmu}) with $\alpha'=1$.
We note that the discrepancy is not large and is even smaller
for the higher Mott lobes. Indeed  $\bar{U}(T_{\rm c} \to 0) = 2(2\alpha'+1)$ versus
$\bar{U}_{\rm c} = (2\alpha'+1) + \sqrt{(2\alpha'+1)^2 - 1}$. Another 
feature is the maximum in the $\bar{T}_{\rm c}(U)$ curve
(cf. Fig. \ref{qualphasediag} and \cite{Sheshadri1993}).
Both features mentioned are caused by the fact that the
two conditions Eqs. (\ref{eq:n012}) and (\ref{eq:n122}) are strictly enforced, whereas
they become less appropriate for small $U$. 
The exact solution \cite{noota} for four slave bosons on a four site lattice for 
small $\bar{U}$  shows that a better result may
be obtained if a fourth boson occupation number $n^3$ is included in our approach. The set of
equations to be solved then becomes, again for $n=1$,
\begin{eqnarray}
&& n^0 + n^1 + n^2 + n^3  =  1 \label{eq:n0123} \\
&& n^1 + 2 n^2 + 3 n^3  =  1  \label{eq:n12233} \\
&& \frac{3}{\bar{\mu} - 2\bar{U}}\left( n^3 - n^2 \right) + 
    \frac{2}{\bar{\mu} - \bar{U}}\left( n^2 - n^1 \right) + 
    \frac{1}{\bar{\mu}} \left( n^1 - n^0 \right)  =  1  ~.
\label{eq:critc4}
\end{eqnarray} Again $n^0, n^2$, and $n^3$ can easily be expressed in terms of $n^1$, but
no exact solution appears to be possible in this case. However, we have 
managed to find solutions numerically. The results
for $T_{\rm c}$ are depicted in Fig. \ref{4bosons.eps} and show fairly little quantitative
change compared to the analytical result Eq. (\ref{eq:Tc3b1}). In particular,
$\bar{T}_{\rm c}$ still vanishes for $\bar{U} \approx 6$, and the maximum is still
there, although shifted to a lower $\bar{U} \approx 1.8$ compared to $\bar{U}=2.15$
for Eq. (\ref{eq:Tc3b1}). It is satisfactory to find that for the higher values
of $\bar{U}$, $n^1$ starts to increase rapidly towards 1, signalling the approach
of the Mott-insulator phase, whereas $n^3$ is almost negligible ($< 1 \%$) already for 
$\bar{U} \approx 3$, supporting a description in terms of 3 slave bosons only \cite{nootb}.

\subsection{Quasiparticle-quasihole dispersion relations} 

\begin{figure}
\includegraphics[width=10cm]{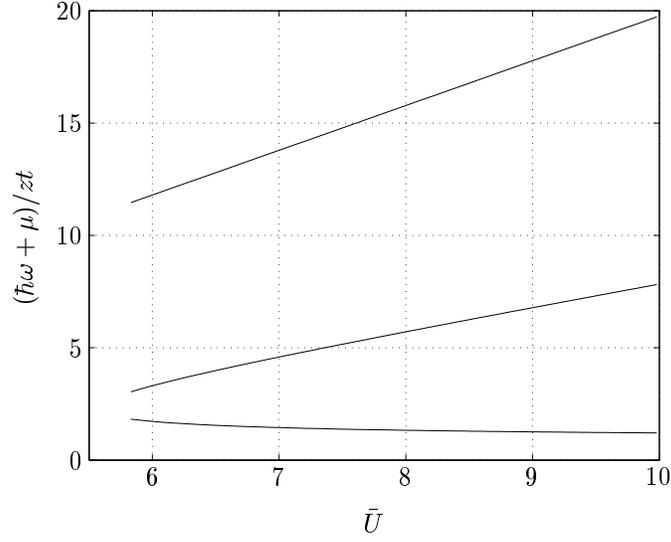}
\caption{The dispersion relations for ${\bf k}=0$ in the case where we take into account higher filling factors
	at nonzero temperature. On the vertical axis is $(\hbar \omega + \mu )/zt$ and on the horizontal axis is $\bar{U}$.
	Here we have taken into account all the terms with $\alpha = 0,1,2$ at a temperature of $z t \beta = 10$. 
}\label{fulldisp}
\end{figure}

\begin{figure}
\includegraphics[width=10cm]{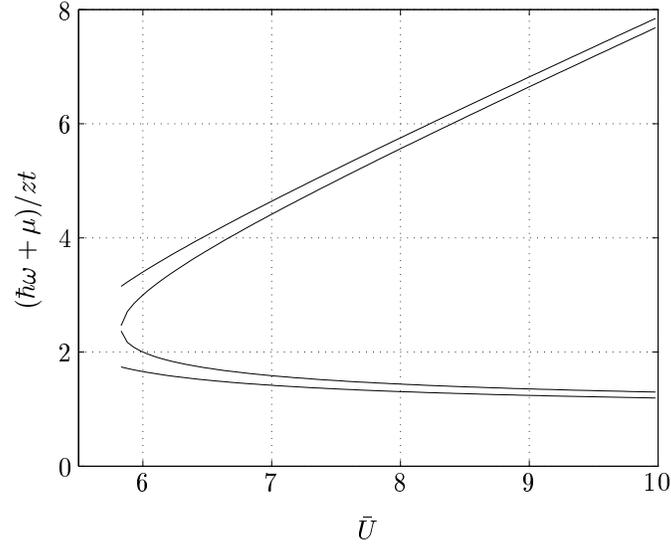}
\caption{Dispersion relations  $\hbar \omega + \mu$ as a function of  $U/zt$ for ${\bf k=0}$ 
for zero and nonzero temperatures. The inner lobe corresponds to zero temperature. The 
outer lobe correponds to a temperature of $z t \beta = 3$. 
Here we have only taken into account the 
first three terms in the right-hand side of  Eq. (\ref{groenefunctie}), i.e., in the sum we only include 
the terms with $\alpha=0$ and $\alpha=1$.
}\label{2termapprox}
 \end{figure}

\begin{figure}[htb!]
\includegraphics[width=10cm]{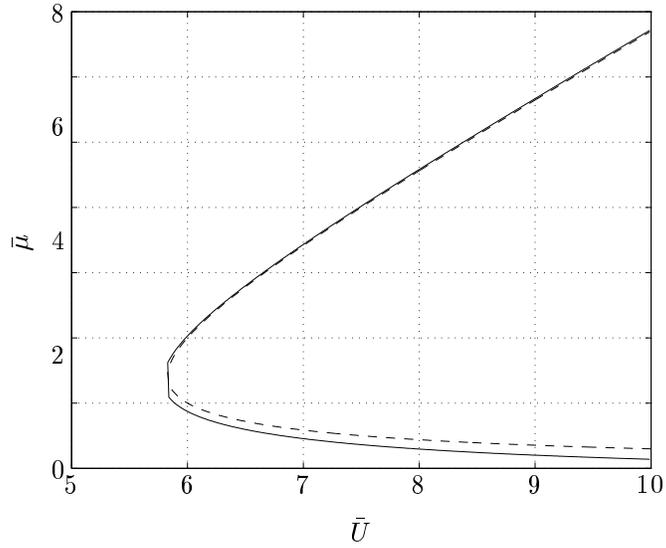}
\caption{The $\bar{\mu}$-$\bar{U}$ phase diagram for zero and nonzero temperatures.
The inner lobe corresponds to the zero-temperature case. The outer lobe corresponds to a temperature of 
$z t \beta = 2$ 
}\label{sfnft}
\end{figure}

Consider now the propagator $G^{-1}({\bf k},\omega)$, given by
\begin{equation}\label{propagator}
-\hbar G^{-1}({\bf k},\omega) = 
\left(
\epsilon_{{\bf k}} - \epsilon_{{\bf k}}^2 
 \sum_{\alpha } (\alpha +1) 
\frac{n^{\alpha} - n^{\alpha + 1}}{-  \hbar \omega  - \mu + \alpha U}
\right).
\end{equation} At zero temperature and for a given integer filling factor $\alpha'$, we have in a mean-field 
approximation that $n^{\alpha} = 
\delta_{\alpha , \alpha'}$ and we retrieve the previously found result for the 
quasiparticle-quasihole dispersions \cite{Oosten2001}. In this case the real solutions of $\hbar \omega$ follow from a 
quadratic equation 
$G^{-1}({\bf k},\omega_{n}) = 0$. At nonzero temperature the occupation numbers in 
general are all nonzero and there will be more than 
just two solutions for $\hbar \omega$. In the set of solutions there are still two solutions that correspond to 
the original single quasiparticle and quasihole dispersions.
The physical interpretation of the other solutions is that they correspond to the excitation of a higher number of 
quasiparticles and quasiholes. In Fig. \ref{fulldisp}. we show the three low lying excitation energies for 
${\bf k}=0$  at a temperature of $z t \beta = 10$.
To obtain an analytic expression for the single quasiparticle-quasihole dispersion 
we only take into account the two terms in the  sum in Eq.(\ref{groenefunctie}) which have numerators $n^{\alpha' -1} - n^{\alpha'}$ and 
$n^{\alpha'} - n^{\alpha' + 1}$. These correspond to processes where the occupation of 
a site  changes between $\alpha' - 1 , \alpha'$ and $\alpha' + 1$. We find

\begin{eqnarray}\label{singleqpqhdisp}
&& \hbar \omega^{qp,qh}_{\bf k} =
- \mu + \frac{U}{2}  + \frac{1}{2} \epsilon_{{\bf k}} 
(\alpha' n^{\alpha' -1} - n^{\alpha'} + (\alpha' + 1) n^{\alpha' +1})
\nonumber \\ &&  \pm \frac{1}{2} 
\sqrt{
U^2 + 2 (\alpha' n^{\alpha' -1} - (1 + 2 \alpha') n^{\alpha'} + 
(1 + \alpha')  n^{\alpha'+1}) U 
\epsilon_{{\bf k}} + (
\alpha n^{\alpha' -1 } + n^{\alpha'} - (1+\alpha') n^{\alpha'+1})^2 
\epsilon_{{\bf k}}^2 .
}  \nonumber \\
\end{eqnarray} In Fig. \ref{2termapprox} we have plotted these dispersions 
at ${\bf k}=0$ as a function of $U$. Comparison with 
Fig. \ref{fulldisp} shows that Eq. (\ref{singleqpqhdisp}) gives an appropriate description of the 
single quasiparticle-quasihole dispersions.
As can be seen from Fig. \ref{2termapprox}  the tip of the lobe moves to smaller $U$
as a function of increasing temperature. This can be understood because that 
point now describes the superfluid-normal phase transition (cf. Figs. \ref{qualphasediag}, 
\ref{4bosons.eps}). In Fig. \ref{sfnft} we show how the superfluid-normal boundary in the  $\bar{\mu} - 
\bar{U}$ plane evolves for nonzero temperatures.
If we define the gap as the difference between the two solutions 
at ${\bf k}=0$, we find that the gap grows bigger as the temperature 
increases. 
As we have seen in Sec. \ref{sec3b} it is incorrect, however, to conclude from this 
that the region of the 
Mott insulating phase in the $\mu$-$U$ phase diagram grows as temperature increases.
As mentioned previously, strictly speaking there is no Mott insulator away from 
zero temperature and at nonzero temperatures there is only a crossover between 
a phase which has a very small compressibillity and the normal phase.

\section{Fluctuations}\label{fluctuations}

\begin{figure}[htb!]
\includegraphics[width=10cm]{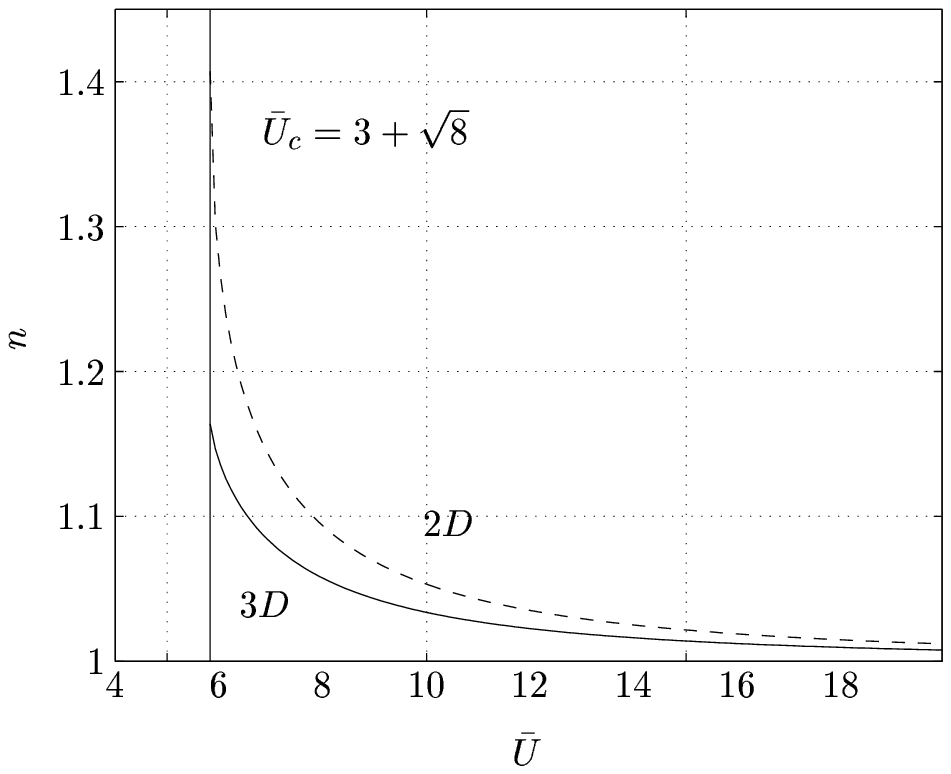}
\caption{Total density $n$ as a function of interaction strength 
$U/zt$ for the first Mott lobe in two and three dimensions when including fluctuations.
The density approaches a finite value different from one, when approaching $U_{c}$.}\label{123ddens}
\end{figure}

In this section we make a first step towards the study of fluctuation effects and 
derive an identity between the atomic Green's function  and the superfluid Green's function
in Eq. (\ref{groenefunctie}). This we then use to calculate the atomic particle density.
In appendix \ref{subtleties} we show that the easiest way to calculate the density is by
 making use of currents that couple to the atomic fields. We start with the action of the Bose-Hubbard model
\begin{equation}
S[a^{*},a] = \int_{0}^{\hbar \beta} d \tau  \left[ 
\sum_{i} a^{*}_{i} \left( \hbar \frac{\partial}{\partial \tau} - \mu \right) a_{i}
- \sum_{ij} t_{ij} a^{*}_{i} a_{j} + \frac{U}{2}  \sum_{i} a^{*}_{i} a^{*}_{i} a^{\phantom * }_{i} 
a^{\phantom *}_{i}
\right] .
\end{equation} We are interested in calculating the $\langle a^{*}_{i} a_{i} \rangle$ correlation function.
Therefore we add currents $J^{*}, J$ that couple to the $a^{*}$ and $a$ fields as
\begin{equation}
Z[J^{*},J] = \int d[a^{*}] d[a] \exp{\left\{
-S_{0}/\hbar + \frac{1}{\hbar} \int_{0}^{\hbar \beta} d \tau \sum_{ij} a^{*}_{i} t_{ij} a_{j} + \int_{0}^{\hbar \beta} 
d \tau \sum_{i} 
\left[ 
J_{i}^{*} a_{i} + a_{i}^{*} J_{i}\right]
\right\}}.
\end{equation} 
Here  $S_{0} = S_{0}[a^{*}, a]$ denotes the action for $t_{ij} = 0$. 
The most important step in the remainder of the calculation is to a perform again a Hubbard-Stratonovich 
transformation by adding a complete square 
to the action. The latter can be written as,
\begin{equation} \int d \tau
\sum_{i,j} \left(
a^{*}_{i} - \Phi^{*}_{i} + \hbar t^{-1}_{ij'} J^{*}_{j'} 
\right)
t_{ij}
\left( 
a_{j} - \Phi_{j} + \hbar t^{-1}_{jj''} J_{j''} 
\right).
\end{equation} Straightforward algebra yields
\begin{equation}
Z[J^{*},J] = \int d[\Phi^{*} ] d[\Phi] \exp{\left\{
\sum_{{\bf k},n} \left( -\hbar
\Phi^{*}_{{\bf k},n} G^{-1}({\bf k},i \omega_{n}) \Phi_{{\bf k},n} + J^{*}_{{\bf k},n} \Phi_{{\bf k},n} + 
J_{{\bf k},n} \Phi^{*}_{{\bf k},n} - \frac{\hbar}{\epsilon_{{\bf k}}} J^{*}_{{\bf k},n} J_{{\bf k},n} \right) 
\right\}}.
\end{equation} Differentiating twice with respect to the currents gives then the relation
\begin{equation}
\left. \frac{1}{Z[0,0]} \frac{\delta^2}{\delta J^{*}_{{\bf k},n} \delta J_{{\bf k},n}}  Z[J^{*},J] \right|_{J^{*},J = 
0 } = \langle a_{{\bf k},n}^{*} a_{{\bf k},n} \rangle = \langle \Phi_{{\bf k},n}^{*} \Phi_{{\bf k},n} \rangle - 
\frac{\hbar}{\epsilon_{{\bf k}} }.
\end{equation} This is very useful indeed since the  correlator 
$ \langle \Phi_{{\bf k},n}^{*} \Phi_{{\bf k},n} \rangle = - G({\bf k},i \omega_{n}) $.
At zero temperature the retarded Green's function can be written as
 \begin{subequations}
\begin{equation}\label{greenfn}
-\frac{1}{\hbar} G({\bf k},\omega) = \frac{Z_{\bf k}}{- \hbar \omega + \epsilon^{qp}_{\bf k}} + \frac{1-Z_{\bf 
k}}{-\hbar 
\omega + 
\epsilon^{qh}_{\bf k}} + \frac{1}{\epsilon_{{\bf k}}},
\end{equation} where the wavefunction renormalization factor is 
\begin{equation}
Z_{\bf k} =  
\frac{U(1 + 2 \alpha') - \epsilon_{{\bf k}} + 
\sqrt{U^2 - 2 U \epsilon_{{\bf k}} (1 + 2 \alpha') + \epsilon_{{\bf k}}^{2}}}
{2 \sqrt{U^2 - 2 U \epsilon_{{\bf k}} (1 + 2 \alpha') + \epsilon_{{\bf k}}^{2}}},
\end{equation} and 
\begin{equation}
\epsilon^{qp,qh}_{\bf k}  = - \mu + \frac{U}{2} ( 2 \alpha' - 1) - 
\frac{\epsilon_{{\bf k}}}{2} \pm \frac{1}{2} \sqrt{\epsilon_{{\bf k}}^2 - (4 \alpha' + 
2) U \epsilon_{{\bf k}} + U^2} .\end{equation}

\end{subequations}
Note that 
$Z_{\bf k}$ is always positive and  in the limit where $U \rightarrow \infty $ we 
have that $Z_{\bf k} \rightarrow (1 + \alpha')$. The quasiparticle dispersion
$\epsilon^{qp}_{\bf k}$ is always greater than or equal to zero and 
$\epsilon^{qh}_{\bf k}$ is always smaller than or equal to zero. Because of this 
only the quasiholes give a contribution to the total density at zero temperature.
The density can be calculated from,
\begin{equation}
n = \frac{1}{N_{s} \hbar \beta} \sum_{{\bf k},n} \langle a_{{\bf k},n}^{*} a_{{\bf k},n} \rangle = 
\frac{1}{N_{s} \hbar \beta} \sum_{{\bf k},n} \left\{
-\hbar G({\bf k},\omega_n) - \frac{\hbar}{\epsilon_{{\bf k}}} \right\} \overset{\footnotesize{\beta 
\rightarrow 
\infty}}{=} \frac{1}{N_{s}} \sum_{{\bf k}} ( Z_{\bf k} - 1 )
\overset{\footnotesize{U \rightarrow \infty}}{=} \alpha'.
\end{equation} If we expand the square-root denominator of $Z$ for small ${\bf k}$ we see that it behaves as 
$ 1/{\bf k}$, therefore in two and three dimensions we expect the integration over ${\bf k}$ to converge.
In \mbox{Fig. \ref{123ddens}} we have plotted the density for $\alpha' = 1$ as given by the equation above.
We see that the density quickly converges to one, but near the tip of the 
Mott lobe in all dimensions it deviates significantly  from one. 
This result is somewhat unexpected and may be due to the break-down of the gaussian approximation
near the quantum phase transition. A more detailed study of the fluctuations is
beyond the scope of the present paper and is therefore left to future work.

\section{Conclusions}

In summary, we have applied the slave-boson formalism to the Bose-Hubbard model, which enabled us to 
analytically describe the physics of this model at nonzero temperatures.
We have reproduced the known zero-temperature results and we have computed
the critical temperature for the  superfluid-normal phase  transition. 
The crossover from a Mott insulator to a normal phase  has also been 
quantified. We have shown how  thermal  fluctuations introduce additional dispersion modes associated with paired
quasiparticles-quasiholes propagating  through the system. 
 We have also considered density fluctuations induced by the creation
of quasiparticle-quasihole pairs. These fluctuations do not average out to zero
in the gaussian approximation.

\begin{acknowledgments}
This work is part of the research programme of the `Stichting voor Fundamenteel Onderzoek der
Materie (FOM)', which is financially supported by the `Nederlandse Organisatie voor
Wetenschaplijk Onderzoek (NWO)'.
\end{acknowledgments}

\appendix
\section{Higher-order terms}\label{higherorder}

If we also want to calculate quantities like the superfluid density, we have to calculate the effective 
action up to  fourth order. One way to do this is by going to higher order in the interaction part. 
Here we follow a  slightly different strategy. Because we are only interested in the mean-field theory, 
it suffices to just consider  $\Phi_{{\bf 0},0}$ terms.  The effective action for $\Phi_{{\bf 0},0}$ is found from
\begin{equation}
Z = \int \frac{d [\Phi^{*}_{{\bf 0},0} ] d [\Phi_{{\bf 0},0} ] }{\hbar \beta} 
\int \prod_{\alpha, {\bf k},n}\frac{d [(a^{\alpha})^{*}_{{\bf k},n}] d[a^{\alpha}_{{\bf k},n}]}{\hbar \beta}
\exp{\left( -\frac{1}{\hbar} S \right)},
\end{equation} where  from Eq. (\ref{3133}) we have
\begin{equation}
S = i N_{s} \hbar \beta \lambda + \epsilon_{\bf 0} | \Phi_{{\bf 0},0} |^2   + \sum_{\alpha \beta}
\sum_{{\bf k},n} 
(a^{\alpha}_{{\bf k},n})^{*} 
M^{\alpha \beta}
a^{\beta}_{{\bf k},n}.
\end{equation} Note, however, that now the matrix $M$ is only blockdiagonal and it contains off-diagonal terms 
proportional to 
$\Phi_{{\bf 0},0}$. When we take the determinant of that matrix, you get automatically all powers in 
$\Phi_{{\bf 0},0}$. This can be made more explicit by looking at the block-structure of the matrix which is
\begin{subequations}
\begin{equation}
M =
\left(
\begin{matrix}
B_{0}  \\
& B_{2} \\
& & B_{4} \\
& & & \ldots \\
\end{matrix}
\right),
\end{equation} where
\begin{equation}
B_{\alpha} = 
\left(
\begin{matrix}
\chi_{\alpha} & \frac{\sqrt{\alpha +1}}{\sqrt{N_{s} \hbar \beta}} \epsilon_{\bf 0} \Phi_{{\bf 0},0} \\
\frac{\sqrt{\alpha +1}}{\sqrt{N_{s} \hbar \beta}} \epsilon_{\bf 0} \Phi_{{\bf 0},0}^{*} & \chi_{\alpha + 1}  \\
\end{matrix} \right),
\end{equation} with $\chi_{\alpha} =  - i \hbar \omega_{n} - i  \lambda  -\alpha \mu + 
 \alpha (\alpha -1) U/2 $.
\end{subequations} 
The slave bosons can be integrated out with the result
\begin{equation}\label{hmmm}
Z = \int \frac{d [\Phi^{*}_{{\bf 0},0} ] d [\Phi_{{\bf 0},0} ] }{\hbar \beta} 
\exp{
\left\{ -\frac{1}{\hbar} \left( i N_{s} \hbar \beta \lambda + \epsilon_{\bf 0} | \Phi_{{\bf 0},0} |^2   \right) \right\} }
\exp{\left\{ -\sum_{{\bf k},n} \log{ \left[ {\rm det} \beta M \right]} \right\}} .	
\end{equation}

The determinant can be calculated up to fourth order in $\Phi_{{\bf 0},0}$ as
\begin{equation}
\det \beta M = \left(
\prod_{\alpha} \beta \chi_{\alpha} \right) \left( 
1 
+
\sum_{\alpha }
\frac{\epsilon_{\bf 0}^2}{N_{s} \hbar \beta} | \Phi_{{\bf 0},0} |^2  \frac{(\alpha + 1)}{\chi_{\alpha} \chi_{\alpha+1}}
+ 
\sum_{\alpha}\sum_{|\alpha - \beta| \geq 2 } 
\frac{\epsilon_{\bf 0}^4}{(N_{s} \hbar \beta)^4} | \Phi_{{\bf 0},0} |^4 
 \frac{(\alpha+1) (\beta + 1)
}{\chi_{\alpha} \chi_{\alpha+1} \chi_{\beta} \chi_{\beta+1}}
\right).
\end{equation} For small $\Phi_{{\bf 0},0}$ we can expand the logarithm in Eq. (\ref{hmmm}) by using
the Taylor expansion 

\[
\log{\left\{ 1 - \alpha x^2 + \gamma x^4 \right\} } = - \alpha x^2 + 1/4
(-2 \alpha^2 + 4 \gamma) x^4 + \mathcal{O}(x^5)  .
\] 
Combining the latter equation with  Eq. (\ref{hmmm}), we also recover that the second-order term in the effective action 
for  $\Phi_{{\bf 0},0}$ is given by
\begin{equation}\left( \epsilon_{\bf 0}
- \hbar \sum_{{\bf k}, n } \sum_{\alpha }
\frac{\epsilon_{\bf 0}^2}{N_{s} \hbar \beta}   \frac{(\alpha + 1)}{\chi_{\alpha} \chi_{\alpha+1}} \right) |\Phi_{{\bf 0},0}|^2
=  \left( \epsilon_{\bf 0} + \epsilon_{\bf 0}^{2} \frac{n^{\alpha} - n^{\alpha + 1}}{-\mu + \alpha U}\right) |\Phi_{{\bf 0},0}|^2
= -\hbar G^{-1}({\bf 0},0) |\Phi_{{\bf 0},0}|^2.
\end{equation} 

We determine the effective action to fourth order in the case of 
the first four slave bosons. Using the above we  can readily 
verify that
\begin{eqnarray}\label{nofootbaltoday}
&& - S^{\rm eff}/\hbar =
 - \frac{1}{\hbar} \left( \epsilon_{\bf 0} |\Phi_{{\bf 0},0}|^2 + i N_{s} \hbar \beta \lambda  
  - \sum_{j=0}^{3} \log{\beta \chi_{j}}
\right.  \nonumber \\ && \left.
- \log{
\left(
 1  -
 (\frac{\epsilon_{\bf 0}}{N \hbar \beta} )^2 
\left( \frac{3}{ \chi_{3} \chi_{2}} + \frac{2}{ \chi_{2}\chi_{1}} + \frac{1}{\chi_{1} \chi_{0}}
\right) 
 |\Phi_{{\bf 0},0}|^2
 + (\frac{ \epsilon_{\bf 0}}{N_{s} \hbar \beta} )^4  \frac{3}{\chi_{0} \chi_{1} \chi_{2} \chi_{3} }
|\Phi_{{\bf 0},0}|^4
\right) } \right) .
 \nonumber \\
\end{eqnarray} From this we find that $a_{4}$ in the  case of four slave bosons is given by
\begin{equation} a_{4} = 	
\frac{\hbar}{4} \left(\frac{\epsilon_{\bf 0}}{\sqrt{N_{s} \hbar \beta}}\right)^4 
 \sum_{{\bf k},n}
\left(
- 2  \left( \sum_{\alpha=0}^{3} \frac{(\alpha + 1)}{\chi_{\alpha} \chi_{\alpha+1}} \right)^2 
+  \frac{12}{\chi_{0} \chi_{1} \chi_{2} \chi_{3}}
\right),
\end{equation} or explicitly,
\begin{eqnarray}\label{complicatedexpression}
a_4 &=& - \left( \frac{\epsilon_{\bf 0}}{2 N_{s}^{2} \hbar \beta} \right) 
 \left\{
\frac{9}{(2 \bar{U} - \bar{\mu})^2 }
\left( 3 n^{3} (1 - n^{3}) + 2 n^{2} (1 - n^{2})
 \right)
+
 \frac{18}{( 2 \bar{U} - \bar{\mu})^3 }
 \left( n^{3} - n^{2} \right)
 \right.
\nonumber \\
&& \left.
+
\frac{4}{(\bar{U} - \bar{\mu})^2} \left(
2 n^{2} (1 - n^{2})
+ n^{1} (1 - n^{1})
\right)   +
\frac{8}{(\bar{U} - \bar{\mu})^3} \left(
n^{2} - n^{1}
\right)
 \right.
\nonumber \\
&& \left.
+
\frac{1}{(\bar{\mu})^2 } \left( n^{0} (1 - n^{0}) + n^{1} (1 - n^{1})
\right)
  + \frac{2}{\bar{\mu}^3} \left( n^{0} - n^{1} \right) +
\frac{4}{(\bar{U} - 2 \bar{\mu}) \bar{\mu}^2  }   n^{0} -
\frac{4}{(\bar{U} - 2 \bar{\mu}) (\bar{U} - \bar{\mu})^2  }   n^{2} 
 \right.
\nonumber \\
&& \left.
+
\frac{4}{(\bar{U} - \bar{\mu}) \bar{\mu}  }   n^{1}(1 - n^{1}) 
-
\frac{4 \bar{U}}{(\bar{U} -  \bar{\mu})^2 \bar{\mu}^2  }   n^{1}
-\frac{12}{ (3 \bar{U} - 2 \bar{\mu}) ( 2 \bar{U} - \bar{\mu})^2 } n^{3}  
 \right. \nonumber \\ && \left. -\frac{12}{(2 \bar{U}^2 - 3 \bar{U} \bar{\mu} + \bar{\mu}^2) } 2 n^{2} ( 1 - n^{2} ) -\frac{12
\bar{U}}{(2 \bar{U}^2 - 3 \bar{U} \bar{\mu} + \bar{\mu}^2)^2 } n^{2} + \frac{12}{ (3 \bar{U} - 2 \bar{\mu}) ( \bar{U} -
\bar{\mu})^2 } n^{1} \right\}.  \nonumber \\ \end{eqnarray} Note that in the zero-temperature limit for the first Mott lobe, when the 
slave-boson occupation numbers are proportional to a Kronecker delta,
this result coincides exactly with the one previously derived in standard perturbation theory (cf. Ref. \cite{Oosten2001} ).

\section{density calculations}\label{subtleties}

In this section we demonstrate for the noninteracting case the equivalence of the calculation of 
the total particle density through the thermodynamic relation $N = - \partial \Omega / \partial \mu $
and through the use of source currents that couple to the atomic fields.
We consider a system of noninteracting bosons described by creation and annihilation fields 
$a^{*}_{i}(\tau) $ and $a_{i}(\tau) $ on a lattice. First we calculate the generating functional $Z[J^{*},J]$
for this system,
\begin{equation}
Z[J^{*},J] = \int d[a^{*}] d[a]  \exp{ \left\{ -\frac{1}{\hbar} 
S_{0}[a^{*},a] + 
 \frac{1}{\hbar} \int d \tau 
\sum_{i j } a^{*}_{i} t_{ij} a_{j} + \int d \tau \sum_{i} \left( J^{*}_{i} a_{i} + a_{i}^{*} J_{i} \right) 
\right\} }.
\end{equation} In this equation $S_{0}$ is the on-site action, which in frequency-momentum representation 
typically looks like
\begin{equation}
S_{0} [a^{*},a] = 
\sum_{{\bf k},n} a^{*}_{{\bf k},n} \left( - i \hbar \omega_{n}  - \mu \right) a_{{\bf k},n}.
\end{equation} The hopping term is decoupled by means of a Hubbard-Stratonovich transformation, i.e., we add 
the following complete square to the action,
\begin{equation*}
\sum_{ij} 
\left( a_{i}^{*} - \Phi_{i}^{*} + \hbar \sum_{j'} t^{-1}_{i j'} J^{*}_{j'} \right)
t_{ij}
\left( a_{j} - \Phi_{j} + \hbar \sum_{j''} t^{-1}_{i j''} J_{j''} \right).
\end{equation*}

The atomic fields $a^{*}$, $a$ can now be integrated out. Going through the 
straightforward algebra one arrives at the following expression for the generating functional,
\begin{equation}
Z[J^{*},J] = \int d[\Phi^{*}] d[\Phi] \exp{\left\{
\sum_{{\bf k},n}  \Phi^{*}_{{\bf k},n}  G^{-1}({\bf k}, i \omega_{n}) \Phi_{{\bf k},n} + J^{*}_{{\bf k},n} 
\Phi_{{\bf k},n} 
+ J_{{\bf k},n} \Phi^{*}_{{\bf k},n} 
- \hbar \frac{J_{{\bf k},n} J^{*}_{{\bf k},n} }{\epsilon_{{\bf k}}}
 \right\}},
\end{equation} where $-\hbar G^{-1}({\bf k},i \omega_{n}) = \epsilon_{\bf k} - \epsilon_{\bf k}^{2} \left( 
-i \hbar \omega_{n}  - 
\mu\right)^{-1} $.
The total density may be calculated from this expression by first calculating the correlator $\langle a^{*}_{{\bf k},n} 
a_{{\bf k},n} \rangle$ through functional differentiation with respect to the source-currents $J$, and then to sum over all 
momenta and Matsubara frequencies.  We have for the first step
\begin{equation}
\langle a^{*}_{{\bf k},n} a_{{\bf k},n} \rangle 
= \frac{1}{Z[0,0]}  \left. \frac{\delta^{2}}{\delta J^{*}_{{\bf k},n} \delta J_{{\bf k},n}} Z[J^{*},J] \right|_{J^{*},J=0}
= \frac{\hbar}{-i\hbar \omega_{n} - \mu - \epsilon_{\bf k}}.
\end{equation} We see that there is a pole here at $i \hbar \omega_{n} = -\epsilon_{\bf k} - \mu$.
The density now can be calculated from 
$n = (1/ N_{s} \hbar \beta) \sum_{{\bf k},n} \langle a^{*}_{{\bf k},n} a_{{\bf k},n} \rangle $.
This is the expected result.

On the other hand, we can also calculate the density from the thermodynamic potential $\Omega$, by using the 
relation $N = -\partial \Omega / \partial \mu$ where $N$ is the total number of particles. Doing that for this case we use that 
\begin{equation}
\Omega = \frac{1}{\beta} \sum_{{\bf k},n} \left\{ \log{\left[\beta( - i \hbar \omega_{n} - \mu) \right]}
+ \log{\left[ - \hbar \beta G^{-1}({\bf k},i \omega_{n}) \right] }\right\}
\end{equation} and obtain
\begin{eqnarray}
n &=& - \frac{1}{N_{s} } \frac{\partial \Omega}{\partial \mu} 
 = \frac{1}{N_{s} \hbar \beta}
 \sum_{{\bf k},n} \left\{ 
\frac{\hbar}{- i \hbar \omega_{n} - \mu} +
\frac{\hbar}{- i \hbar \omega_{n} - \mu 
- \epsilon_{\bf k}} \cdot \frac{\epsilon_{\bf k}}{-i \hbar \omega_{n}  - \mu}
\right\}.
\end{eqnarray} When doing the sum over Matsubara frequencies
the pole at $i \hbar \omega_{n} = - \mu$ in the 
first term in the right-hand side is canceled by the second term
and only the other pole at $i \hbar \omega_{n} = -\epsilon_{\bf k} - \mu$ gives a contribution.
This shows the equivalence of both methods.

\newpage

\newpage

\newpage

\newpage

\begin{thebibliography}{16}
\bibitem[{rem()}]{remedy}
\bibinfo{note}{A remedy for this discrepancy may be envisioned in allowing the
  right-hand side of Eq. (\ref{eq:n122}) to deviate sligthly
  from 1. We do not explore this possibility here.}
\bibitem[{n3m()}]{n3mn2}
\bibinfo{note}{We have also studied the case where the ($n^3 - n^2$)-term is
  included with $n^3$ put equal to 0. This leads to the undesirable feature
  that $T_{\rm c} \to 0$ for $U \to 0$. Including this term, that is
  associated with processes involving triply occupied sites, probably is inconsistent
  with allowing a maximum occupancy of two.}
\bibitem[{noota()}]{noota}
\bibinfo{note}{
For instance, at $U=0$, the
3-slave-boson approach can only result in $n^0 = n^1 = n^2 = 1/3$, 
whereas the exact solution for 4 bosons on a 4-site lattice gives
$n^0 = 0.35, n^1 = 0.38, n^2 = 0.20$, $n^3 = 0.059$, and $n^4 = 0.008$
at $\beta = 1$ \cite{BatScal}.
}
\bibitem[{nootb()}]{nootb}
\bibinfo{note}{
In particular, for $\bar{U} = 4$ we find at $T_{\rm c}$ that
$n^0 = 0.125, n^1 = 0.751, n^2 = 0.122$, and $n^3 = 0.0018$, whereas the
3-slave-boson result is 
$n^0 = 0.111, n^1 = 0.778$, and $n^2 = 0.111$.
}
\end{thebibliography}
\end{document}